\documentclass[11pt]{article}
\usepackage{moriond,epsfig}

\def\be{\begin{equation}}
\def\ee{\end{equation}}
\def\bea{\begin{eqnarray}}
\def\eea{\end{eqnarray}}

\def\sgr {SGR~1806--20}

\def\xmm  {\textit{XMM-Newton}}
\def\int  {\textit{INTEGRAL}}

\begin{document}
\vspace*{4cm}
\title{The Highest Magnetic Fields in the Universe: \\
Anomalous X-Ray Pulsars and Soft Gamma-ray Repeaters}

\author{ S. MEREGHETTI }

\address{INAF - IASF Milano, via Bassini 15\\
I-20133 Milano, Italy}

\maketitle\abstracts{In the last few years it has been recognized
that two apparently distinct classes of peculiar high-energy
sources are actually related and can be explained as young neutron
stars with magnetic fields as high as 10$^{14}$--10$^{15}$ Gauss.
One of these ''magnetars'', \sgr,  has recently emitted the most
powerful giant flare ever recorded. The high-energy observations
of \sgr\ carried out with \xmm\ and \int\ in the past two years
showed a long term trend of increasing activity preceding the 2004
December 27 event.  \int\ data of this giant flare provided unique
evidence for hard X-ray emission lasting about one hour after the
burst onset, possibly due to the interaction of mildly
relativistic ejecta with the circumstellar medium.}

\section{Introduction}

Observations carried out in the last decade have provided
increasing evidence for the existence of neutron stars with
magnetic fields as high as 10$^{14}$--10$^{15}$ G, or magnetars
\cite{dt92,td95}. These objects, about a dozen of which are
currently known, are powered by magnetic energy, contrary to the
nearly 2000 neutron stars which are observed either as
rotation-powered pulsars or X--ray sources powered by accretion in
binary systems.

Two different classes of objects are thought to be magnetars: the
Soft Gamma-ray Repeaters, discovered as sources of short bursts of
hard X-rays ($>$30 keV) with super-Eddington
luminosity~\cite{la87}, and the Anomalous X--ray Pulsars,
discovered as persistent, soft ($<$10 keV) X--ray sources with
pulsations of several seconds and spinning-down on time scales of
$\sim10^{4}$--10$^{5}$ years~\cite{ms95}. An increasing number of
common properties,  pointing to a close relationship  of these two
apparently different classes of objects and leading to their
interpretation as magnetars, has been found. Several good reviews
\cite{hurleyrew,me02,wt04} are available on these sources. Here I
will focus on some recent results on \sgr\ obtained with  \xmm\
and \int, the two major satellites for high-energy astrophysics of
the European Space Agency.

\section{The persistent emission from \sgr  }

\begin{figure}
\begin{center}
 \psfig{figure=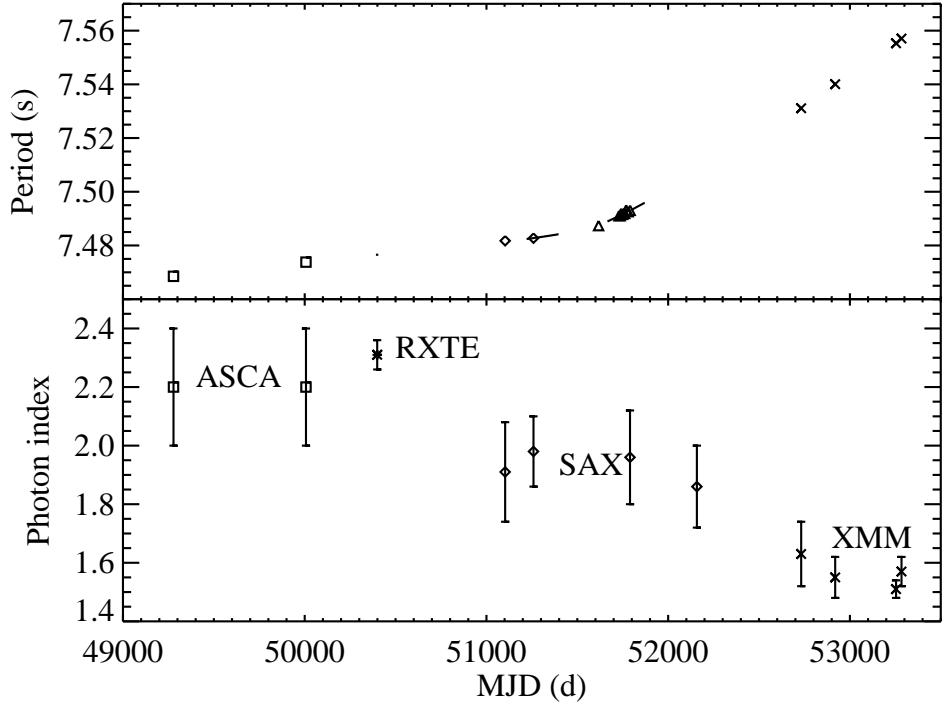,width=10cm,angle=90}
\end{center}
 \caption{Long term evolution of the pulse period
and energy spectrum of SGR 1806--20. The power law photon index
decreased from 2.2 to 1.5, indicating a spectral hardening, while
the average spin-down rate changed from $\sim$8.5$\times10^{-11}$
s s$^{-1}$ to 5.5$\times10^{-10}$ s s$^{-1}$.}
\end{figure}

Four observations of \sgr\ carried out with \xmm\ in 2003-2004
have provided the best  spectroscopic results ever obtained in the
1-10 keV range for this source  and allowed us to study its long
term evolution with a homogeneous set of data from the same
instrument \cite{me05xmm}. The source luminosity in Fall 2004 was
$\sim10^{36}$ erg s$^{-1}$, a factor two higher than that measured
in all the previous observations. The spectrum on 2004 September 6
required the presence of a blackbody of temperature
$kT_{BB}\sim0.8$ keV, in addition to the power law that was found
adequate to fit all the  spectra obtained earlier (which had a
lower statistics). In fact, although not formally required in the
fits, a blackbody component with constant temperature and
luminosity is compatible with all the \xmm\ observations and the
variations in luminosity and spectral shape can be explained  only
by changes in the power law component \cite{me05xmm}.

Comparison of the \xmm\ results  with earlier measurements
indicates a long term hardening of the spectrum, correlated with
an increase in the source average spin-down rate (see Fig. 1).
Such a correlation, previously noticed \cite{mw01} by comparing
different SGRs and Anomalous X-ray Pulsars, is observed here for
the first time  within the same source.

At higher energies, observations with  \int\ showed, for the first
time in a SGR, the presence of persistent emission (i.e. not due
to bursts) extending up to 150 keV \cite{me05int}. This result was
possible thanks to the good sensitivity of the IBIS instrument
\cite{ibis} coupled to its excellent imaging capabilities, which
are essential to disentangle the hard X--ray emission from sources
in crowded Galactic regions. The spectra reported in Fig.~2 show
that the emission in the 20-150 keV range increased and hardened
in Fall 2004, when also a higher rate of emitted bursts was
observed.\cite{me05int} The spectrum of the persistent emission is
harder than that of the bursts (one example is given in Fig.2(c)).
This supports the interpretation of the persistent hard X--ray
emission as a truly different physical component \cite{tlk02}, not
due to the integrated flux of numerous weak bursts too dim to be
seen individually.

Both the \xmm\ and \int\ results indicate a long-term growth in
the level of non-thermal magneto-spherical activity that fits
reasonably well in the scenario of a  magnetar with a
twisted-dipole magnetosphere configuration \cite{tlk02}. In this
model currents flowing in the magnetosphere are expected to lead
to the formation of a high-energy tail  through repeated resonant
scattering of the thermal photons emitted at the star surface. A
gradually increasing twist results in a larger optical depth and
this causes a hardening of the X-ray spectrum. At the same time,
the spin-down rate increases  because, for a fixed dipole field,
the fraction of field lines that open out across the speed of
light cylinder grows. The stresses building up in the neutron star
crust and the magnetic footprints movements can lead to crustal
fractures which can be energetic enough to explain the observed
increase in the bursting activity.

\begin{figure}
\center \psfig{figure=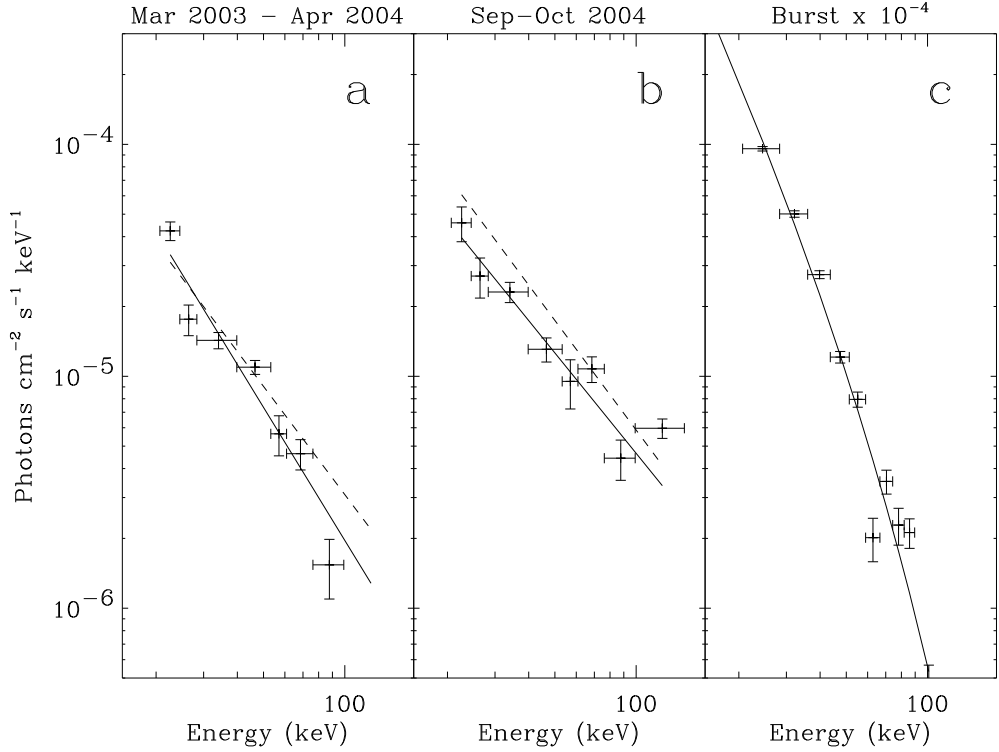,width=10cm,angle=90}
 \caption{IBIS/ISGRI spectra  of  \sgr : (a) persistent emission
March 2003-April 2004, (b) persistent emission Sept.-Oct. 2004,
(c) one burst (scaled down by a factor 10$^{4}$). The solid lines
in (a) and (b) are the best fits power law spectra with photon
index $\Gamma=1.9\pm0.2$ and $\Gamma=1.5\pm0.3$, respectively. The
dashed lines indicate the extrapolation of  power-law spectra
measured in the 1-10 keV range. The solid line in (c) is the best
fit thermal bremsstrahlung spectrum with kT=35 keV.}
\end{figure}

\section{Properties of \sgr\ bursts}

Being at only 10$^{\circ}$ from the Galactic Center direction,
\sgr\ is located in one of the sky regions extensively observed by
\int. Thus  more than 200 bursts have been detected to date,
including the faintest ones ever observed from \sgr\ in the 20-100
keV energy range. This will allow us \cite{gop} to extend the
burst LogN-LogS down to a fluence S$\sim6\times10^{-9}$ erg
cm$^{-2}$.

Time resolved spectroscopy of the bursts with higher fluence
indicates that some of them display significant spectral
variations as a function of time, generally with the softest
emission at the peak \cite{go04}. Analysis of the IBIS data
indicates that the average spectral shape of the bursts  is a
thermal bremsstrahlung with temperature of the order kT$\sim$40
keV (see Fig.~2(c)), consistent with previous measurements of SGRs
bursts in the hard ($>$20 keV) X--ray range. However, such a
spectral shape does not fit well the data below 10 keV, which show
a decrement not compatible with a bremsstrahlung model. This had
already been noticed \cite{fe,ol04} in bursts from SGR 1900$+$14
observed with $BeppoSAX$ and \textit{HETE-2}. Similarly, the
cumulative spectrum (1-10 keV) of the bursts seen in the 2004
\xmm\ observations of \sgr\ are better fitted by a blackbody with
temperature $\sim$2 keV \cite{me05xmm}.

\begin{table}[t]
\caption{Comparison of the three giant flares from SGRs}
\vspace{0.4cm}
\begin{center}
\begin{tabular}{|c|c|c|c|}
\hline
 &  & & \\
Giant Flare  & March 5, 1979 & August 27, 1998 & December 27, 2004 \\
Source   & SGR 0526--66 & SGR 1900+14 & SGR 1806--20 \\
Assumed distance &  55 kpc &  10 kpc &  15 kpc \\
 & & & \\
 \hline
 \textbf{Initial Spike} & & &  \\
 \hline
Duration (s) & $\sim$0.25 & $\sim$0.35 & $\sim$0.5  \\
Peak luminosity (erg s$^{-1}$) & 3.6 10$^{44}$ & $>$3.7 10$^{44}$ & (2$\div5$) 10$^{47}$  \\
 Fluence (erg cm$^{-2}$) & 4.5 10$^{-4}$ & $>$5.5 10$^{-3}$ & 0.6$\div$2  \\
Isotropic Energy (erg)  & 1.6 10$^{44}$ &  $>$6.8 10$^{43}$ & (1.6$\div$5) 10$^{46}$ \\
 & & & \\
 \hline
 \textbf{Pulsating tail} & & & \\
 \hline
Duration (s) & $\sim$200 & $\sim$400 & $\sim$380 \\
 Fluence (erg cm$^{-2}$) & 1 10$^{-3}$ & 4.2 10$^{-3}$  &  5 10$^{-3}$  \\
Isotropic Energy (erg) & 3.6 10$^{44}$ &  5.2 10$^{43}$ & 1.3 10$^{44}$ \\
Spectrum &kT$\sim$30 keV   & kT$\sim$20 keV & kT$\sim$15--30 keV \\
Pulse Period (s) & 8.1  & 5.15  & 7.56  \\
& & & \\
 \hline
  & & & \\
References & \cite{ma99} & \cite{ma99} & \cite{hu05,pa05,te05,ma05}\\
 \hline
\end{tabular}
\end{center}
\end{table}

\section{The giant flare of 2004 December 27}

\begin{figure}
\center \psfig{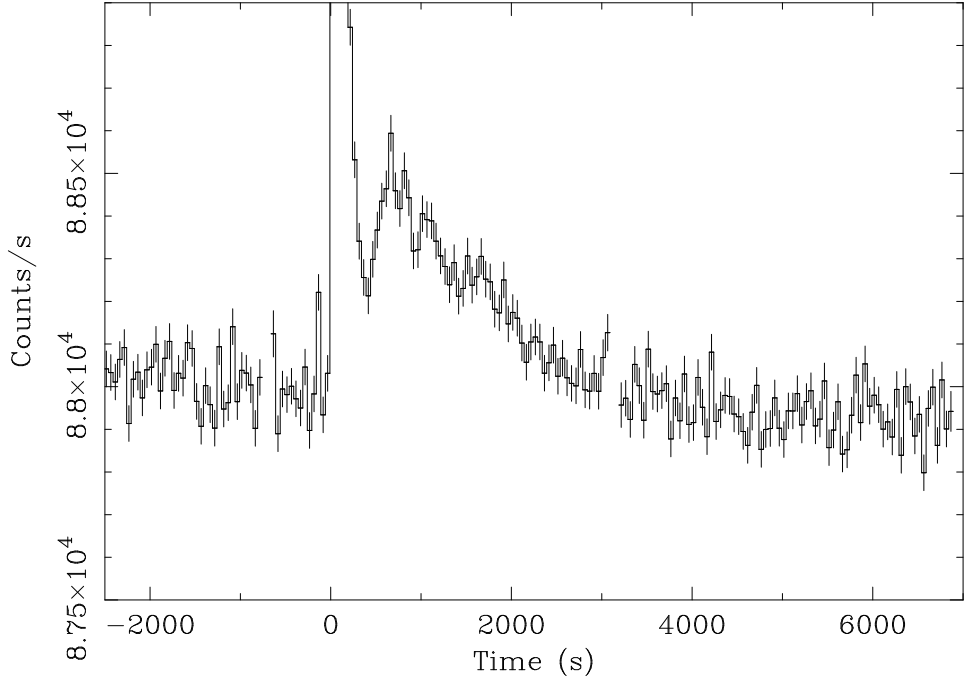}
 \caption{Light curve of the 27 December 2004 giant flare at energy $>$80
keV.  The original data from the SPI-ACS (time resolution of 50
ms) have been rebinned at 50 s to better show the emission lasting
until one hour after the start of the outburst (due to this
rebinning the pulsations at 7.56 s in the time interval 0-400 s
cannot be seen in this plot).  Note that the peak of the flare
reaching an observed count rate $>2\times10^{6}$ counts s$^{-1}$
is out of the vertical scale.}
\end{figure}

The bursting activity of \sgr\ culminated on 2004 December 27,
when the first giant flare from this source was discovered
\cite{bork} with \int\ and detected by more than twenty-one
satellites.\cite{hu05,pa05,te05,ma05,me05acs} This flare produced
the strongest flash of gamma-rays at the Earth ever observed,
causing saturation of most in-flight detectors, significant
ionization of the upper atmosphere \cite{cam05}, and a detectable
flux of radiation backscattered from the Moon.\cite{ma05,me05acs}

Two similar, but less intense, giant flares had been seen in the
past from two other SGRs.\cite{ma79,hu99}   All events had a
similar structure, consisting of a short initial spike with a
harder spectrum, followed by a softer decaying tail lasting
several minutes, modulated at the neutron star rotational period.
The energetics involved in  these three giant flares are compared
in Table 1. Note that, especially for what concerns the initial
spike, the measurements are subject to uncertainties due to
instrumental effects induced by the extremely high photon rate,
which often led to instrument saturation. The range of values
shown in the table for \sgr\ reflects different reports in the
literature. Despite these uncertainties, some interesting
considerations can be done.

For instance it is clear that the energy in the tails of the three
events was roughly of the same order, while the initial spike of
\sgr\ was much more energetic than those of the previous flares
from the other SGRs. The fact that the energy in the tail is
similar in the three giant flares, despite the much higher total
energy release of \sgr\, is consistent with a magnetic field of
the same order in the three sources. In fact the pulsating tail
emission originates from the small fraction of the energy released
during the initial hard pulse that is trapped by closed field
lines in the neutron star magnetosphere forming an optically thick
photon-pair plasma \cite{td95}. The amount of energy that can be
confined in this way is determined by the magnetic field strength,
which is thus inferred to be of the same order in these three
magnetars.

At the time of the flare, \int\ was pointed 106$^{\circ}$ away
from the direction of \sgr\, that therefore was outside the field
of view of its imaging instruments. However, useful data could be
obtained with the Anti-Coincidence Shield (ACS) of the \int\ SPI
instrument. In fact the ACS, with  a total mass of 512 kg of
bismuth germanate  scintillators, besides serving  as a veto for
the SPI germanium spectrometer, works as a sensitive
omnidirectional detector for gamma-ray bursts.\cite{acs} It
provides light curves for photons of energy above $\sim$80 keV in
time bins of 50 ms.  The initial spike of the giant flare from
\sgr\ was so bright to saturate the detector for about 0.6 s and
to cause a reflection from the Moon strong enough to be
detected~\cite{me05acs} by the ACS after a light travel time delay
of 2.8 s.

Contrary to the detectors on other  satellites,  which could
observe the giant flare emission only for a few minutes before it
faded below their sensitivity thresholds, the \int\ ACS detected
an additional component lasting about one hour.\cite{me05acs} As
shown in Fig.~3, this emission is clearly distinct from the
pulsating tail seen in the first $\sim$400 s after the flare
start. It  peaked at t$\sim$650 s and then decayed with time
approximately as a power law $\sim$t$^{-0.85}$. Its fluence above
80 keV was of the order of 3$\times10^{-4}$ erg cm$^{-2}$, which
extrapolating to lower energies and assuming isotropic emission
implies an emitted energy ($>$3 keV) of $\sim1-2\times10^{44}$
ergs. This long lasting emission can be explained as a hard X--ray
afterglow produced by the matter ejected relativistically during
the initial spike. In this case the  bulk Lorentz factor $\Gamma$
can be estimated from the time of the afterglow onset, as in
simple gamma-ray burst afterglow models based on synchrotron
emission \cite{zm04}:\\

$\Gamma\sim$15 (E / 5 10$^{43}$~erg)$^{1/8}$ (n / 0.1
cm$^{-3}$)$^{-1/8}$ (t$_{0}$ / 100 s)$^{-3/8}$ \\

\noindent where $n$ is the ambient density.  The inferred value of
$\Gamma$ is smaller than the typical values of gamma-ray bursts,
but consistent, considering the large involved uncertainties, with
the mildly relativistic outflow derived from the modelling of the
radio data obtained after the \sgr\ flare.\cite{ca05,gr05}



\end{document}